%% file: hls-inv/hls-inv.tex
\documentclass[conference]{IEEEtran}
\IEEEoverridecommandlockouts

\usepackage[colorlinks=false, hidelinks]{hyperref}
\usepackage{balance}
\usepackage{algorithm}
\usepackage{algpseudocode}
\usepackage{amsmath}
\usepackage{enumitem}
\usepackage{graphicx}
\usepackage{subcaption}
\usepackage{listings}
\usepackage{xcolor}
\usepackage{colortbl}
\usepackage{pgfplots}
\usepackage{changepage}
\usepackage{multirow}
\usepackage{url}
\usepackage{tablefootnote}
\usepackage{threeparttable}
\AtBeginDocument{%
  } 

\definecolor{mygood}{RGB}{102, 158, 65}
\definecolor{mybad}{RGB}{242, 98, 105}
\newcommand{\mygood}[1]{\textcolor{mygood}{#1}}

\makeatletter
\newcommand*\myalgsize{%
  \@setfontsize\myalgsize{8}{9}%
}
\makeatother

\begin{document}

\title{AutoINV: Automated Invariant Generation Framework for Formal Verification on High-Level Synthesis Designs}

\author{
\IEEEauthorblockN{
Xiaofeng Zhou$^{*}$, Linfeng Du$^{*}$, Guangyu Hu$^{*}$, Sharad Sinha$^{\dagger}$,
Hongce Zhang$^{\S}$, Wei Zhang$^{*}$
}
\IEEEauthorblockA{
\textit{$^{*}$The Hong Kong University of Science and Technology, Hong Kong}, \textit{$^{\dagger}$Indian Institute of Technology Goa, Goa, India}
}
\IEEEauthorblockA{
\textit{$^{\S}$The Hong Kong University of Science and Technology (Guangzhou), Guangzhou, China}
}
\IEEEauthorblockA{
\{xzhoubu, linfeng.du, ghuae\}@connect.ust.hk, eeweiz@ust.hk, sharad@iitgoa.ac.in, hongcezh@hkust-gz.edu.cn
}
}

\maketitle
\input{hls-inv/abstract.tex}
\input{hls-inv/introduction.tex}

\input{hls-inv/related_work.tex}

\input{hls-inv/Preilimaries.tex}

\input{hls-inv/motivation.tex}
\input{hls-inv/problem_formulation.tex}
\input{hls-inv/method-HLSINV-overall.tex}
\input{hls-inv/method-HLSINV-helper.tex}
\input{hls-inv/method-HLSINV-ranker.tex}
\input{hls-inv/method-HLSINV-prover.tex}

\input{hls-inv/results.tex}
\input{hls-inv/conclusion.tex}



\bibliographystyle{IEEEtran}
\balance
\bibliography{hls-inv.bib}


\end{document}

%% file: hls-inv/abstract.tex
\begin{abstract}
High-level synthesis (HLS) transforms an algorithmic description of hardware from a higher abstraction (e.g., C/C++) into a register-transfer level (RTL) design, offering reduced development time and greater flexibility in design space exploration.
However, such machine-generated RTL designs may contain major functional bugs or security vulnerabilities due to limitations or errors in the HLS tools. One of the most reliable methods to identify these vulnerabilities is formal verification, particularly model checking. Nevertheless, the large size of the generated RTL often causes model checking to struggle to conclude within reasonable time or resource limits.

In this study, we propose utilizing the high-level design features from the HLS flow to construct a set of helper assertions aimed at guiding the model checker and accelerating the verification process. To identify the most effective set of helpers to assist the model checker, we develop a proving mechanism that iteratively reuses proving information to select the potentially most useful set of helpers.

We evaluate the proposed framework on a set of HLS design benchmarks. Experimental results demonstrate that, when compared to vanilla model checking, our approach achieves a speedup of up to 6.05$\times$, and 2.23$\times$ on average.

\end{abstract}

%% file: hls-inv/introduction.tex
\section{Introduction}\label{sec:intro}

\textit{High-level synthesis (HLS)} tools accelerate digital circuit development by compiling functions written in high-level programming languages, such as C and C++, into \textit{register-transfer level (RTL)} descriptions. Although HLS tools have been well-studied in terms of performance optimization and design space exploration (DSE), recent studies~\cite{Herklotz2021, Pundir2022} indicate that commonly used HLS tools are likely to introduce functional and security vulnerabilities into the generated RTL designs. Consequently, there is a growing need to verify the correctness and safety of RTL designs generated by HLS tools.


Hardware formal verification is a viable option for verification. Unlike hardware testing, which provides only limited guarantees depending on the coverage of test cases, hardware formal verification mathematically proves whether a property specification in hardware design holds or not. A common practice in hardware formal verification is SAT-based model checking. SAT-based model checking algorithms, built upon SAT solvers, attempt to establish a proof that guarantees a property specification holds in an RTL design by solving a sequence of SAT problems. These algorithms can produce a proof ensuring that a certain target assertion always holds in the design, regardless of inputs. They can also output a counterexample trace if any states violating the assertion can be reached with a certain sequence of input values. Popular model checking algorithms include the K-induction algorithm, Binary-Decision-Diagram-based model checking, and IC3~\cite{bradley2011sat}, which was later developed into the \textit{Property-Directed Reachability (PDR)} algorithm~\cite{pdr-een2011efficient}. We refer to this commonly used model checking algorithm as the IC3/PDR algorithm in the rest of the paper. In this work, we will perform hardware formal verification on HLS designs using the IC3/PDR algorithm due to its efficiency and potential for optimization and exploration. \looseness=-3

Although model checking is an effective method for detecting potential bugs and producing guarantees, it often encounters scalability issues known as the state space explosion. When the number of registers in the RTL design is substantial or the transition relation becomes complex, model checking algorithms struggle to derive a proof within a reasonable timeframe.\looseness=-1

To alleviate the computational burden of model checking caused by the aforementioned state space explosion problem, one approach is to extract notable facts from the design as \textit{helpers} to aid the verification process. As an example, the study in~\cite{kang2023lfps} identified a set of potential facts of designs and expressed them as RTL assertions. Once proved, these assertions can serve as assumptions that constrain the state space when verifying the property specification under examination. 

The state explosion problem becomes increasingly troublesome, particularly when checking RTL code generated by HLS tools. This code contains not only user-defined software functionalities but also scheduling information produced by HLS tools. Unfortunately, limited efforts have been made to alleviate the computational burden of hardware formal property verification on HLS designs. While it is possible for a verification engineer to manually extract and write helpers to assist the model checking process in small RTL designs, the scale of RTL code in HLS designs is usually large ($\ge$1K lines of Verilog code) and often incomprehensible to human engineers, especially those without hardware expertise.


In this work, we design an automated helper generation framework named AutoINV to assist in model checking of property specifications in HLS designs without any limitations on either design or property specification. There are two main challenges addressed by this framework: first, how to extract high-quality helpers that best assist the verification process of the targeted property specification; and second, once we derive a large set of helpers, how to identify the most effective subset that introduces the least overhead in checking.

To address the first challenge, we design an extraction method that identifies common HLS design features and uses these features to generate helpers. This method is implemented as the AutoINV Helper Generator in the AutoINV framework. For the second challenge, we propose a novel verification method that dynamically selects the most promising helpers based on previous proving trials of the target property. This method is implemented in the AutoINV Helper Ranker module and the AutoINV Prover module.

In summary, this paper makes the following contributions:

\begin{itemize}[noitemsep, topsep=0pt, partopsep=0pt]
     \item We automate the generation of high-quality helpers for formal verification of HLS designs. Specifically, we identify the common hardware patterns in HLS design and derive a template-free helper generation method that takes advantage of these patterns. 
     
     \item We develop an effective helper selection mechanism to identify promising helpers that assist the verification process based on both information from RTL design structure and timeout trials. Unlike black-box approaches such as deep learning models, our method does not require time-consuming data collection and training processes.
     \item We have developed a framework, named AutoINV, for accelerating formal verification tasks in HLS designs, significantly reducing the proving time. Compared to vanilla model checking, our framework achieves an average speedup of 2.23$\times$.

\end{itemize}
 
The remainder of this paper is organized as follows: Section \ref{sec:related_works} reviews related works. Section \ref{sec:pre} presents the background of the IC3/PDR algorithm and the AutoINV framework. Section \ref{sec:motivation} describes the motivation for developing our framework. Section \ref{sec:method} details the functioning of each component in our framework. Section \ref{sec:res} presents the experimental results obtained from a set of benchmarks to demonstrate the effectiveness of our framework. 

%% file: hls-inv/related_work.tex
\section{Related Works}\label{sec:related_works}


\noindent
\textbf{Equivalence checking of HLS-generated designs.}
 Vericert~\cite{FMHLS, FMHLS_inline} is a formally proved HLS tool that generates RTL from C while maintaining functional consistency. The work in~\cite{piccolboni2019kairos} presents KAIROS, a tool to verify the equivalence of RTL implementations generated under different HLS directives. The work SE3~\cite{li2023se3} checks for non-cycle-accurate equivalence between two HLS designs.
These methods are limited to equivalence checking scenarios, while this paper considers accelerating general model checking problems on HLS-generated designs with no restrictions on the property to check. \looseness=-1

\noindent
\textbf{Hardware Property Generation/Mining.}
%
Hardware property mining is a process to mine possible properties from RTL designs. 
For example, IODINE~\cite{iodine05} constructs possible assertions by analyzing simulation traces with fixed patterns;  Goldmine~\cite{2010goldmine, goldmine_memp, goldmine_word} learns assertions from simulation traces. They rank assertions by coverage. The work HARM~\cite{HARM9925689} provides the user with a more flexible assertion generation scheme with a wider range of operators supported.  AssertLLM~\cite{assertLLM} utilizes large language models to generate assertions based on specifications written in natural language.  LFPS~\cite{kang2023lfps} uses RTL templates to generate helper assertions. 

The helper generation in AutoINV can be viewed as hardware property mining, but existing techniques do not directly apply. Prior work mainly mines assertions to improve coverage, whereas we mine helpers that accelerate proving a \emph{given} target property. Moreover, most methods assume human-written RTL; HLS designs are much larger, making the helper search space far bigger and thus expensive to explore with formal guarantees. Finally, many approaches are template-based, but general templates do not capture the recurring patterns in HLS-generated logic effectively. Among prior works,~\cite{xu2023automaticdynopt} is the closest: it generates invariants to speed up reasoning about handshake signals in dynamically scheduled HLS designs. In contrast, our work targets verification-time acceleration for general property checking, rather than design-time optimization for a narrow class of handshake/control properties.\looseness=-1



\noindent
\textbf{Property Ordering/Ranking.} Several studies use statistical information to speed up model checking across many properties. Works such as~\cite{parallel2020FMCAD,parallel20190FMCAD} cluster similar properties using circuit structure and distribute them to different engines based on estimated difficulty. PURSE~\cite{purse24date} further uses IC3/PDR runtime statistics to dynamically schedule property proving. In contrast, our work targets improving the proof of a \emph{single} property and exploits characteristics of HLS-generated designs.


%% file: hls-inv/Preilimaries.tex
\section{Background}\label{sec:pre}
\subsection{IC3/PDR Algorithm}

An RTL design is modeled as a state transition system $\langle V, \mathit{Tr}(V,V'), \mathit{INIT}(V)\rangle$, where $V=\{v_i\}$ are Boolean variables (registers and inputs), $\mathit{INIT}(V)$ characterizes the initial states, and $\mathit{Tr}(V,V')$ is the one-cycle transition relation with $V'$ denoting next-state variables.

Given a safety property $P(V)$, IC3/PDR checks whether a violating state $\neg P(V)$ is reachable (a \emph{bad state}). The algorithm incrementally learns CNF clauses (disjunctions of literals such as $v$ or $\neg v$) to block bad states while maintaining an over-approximation of reachable states; these clauses are also the unit reused by clause side-loading in later sections.

A proof of safety is an inductive invariant $\mathit{INV}(V)$ satisfying Formula~\ref{eq:inductive_invariant}. IC3/PDR constructs such an invariant through a sequence of frames $[F_0,F_1,\ldots,F_N]$, where each $F_i$ is a conjunction of learned clauses and $F_0 \equiv \mathit{INIT}(V)$. The frames are maintained to satisfy Formula~\ref{eq:frame}. The algorithm converges when it finds $N$ such that $F_{N+1}\rightarrow F_N$; together with $F_N\rightarrow F_{N+1}$ implied by the first constraint in Formula~\ref{eq:frame}, this yields a fixed point $F_N \equiv F_{N+1}$, and $F_N$ serves as an inductive invariant proving $P$.

IC3/PDR alternates two subroutines. \textit{Strengthen} searches for counterexamples to the current frames (e.g., a bad state or a state violating the consecution constraint in Formula~\ref{eq:frame}) and refines the frames by adding blocking clauses; if the violation can be traced back to $F_0$, a real counterexample trace is reported. \textit{Propagate} attempts to push learned clauses forward to later frames whenever allowed by Formula~\ref{eq:frame}; successful propagation helps reach the convergence condition.

\begin{equation}\label{eq:inductive_invariant}
\left\{ 
\begin{array}{l}
\mathit{INIT}(V) \rightarrow \mathit{INV}(V) \\
\mathit{INV}(V) \land \mathit{Tr}(V, V') \rightarrow \mathit{INV}(V') \\
\mathit{INV}(V) \rightarrow P(V)
\end{array} 
\right.
\end{equation}

\begin{equation}\label{eq:frame}
\left\{ 
\begin{array}{l}
F_i(V) \rightarrow F_{i+1}(V)\\
F_i(V) \land \mathit{Tr}(V,V') \rightarrow F_{i+1}(V')\\
F_i(V) \rightarrow P(V)
\end{array} 
\right.
\end{equation}

\subsection{Clause Side-loading}

Clause side-loading is a technique implemented in DeepIC3~\cite{hu2024deepic3} to equip the model checking algorithm with a potentially better starting point on $F_1$, where the IC3/PDR algorithm starts. These clauses are checked before loading onto $F_1$ to ensure their correctness. The loaded clauses can provide an initial constraint to the state space and influence the choice of bad states made by the SAT solver, potentially guiding the proof search of the IC3/PDR algorithm into a ``better'' direction. These clauses will also be loaded to later frames automatically by the \textit{propagate} procedure of the IC3/PDR algorithm, forming a stronger guarantee in the proving process of the target property. In DeepIC3, the authors showed the effectiveness of side-loading high-quality clauses by directly picking up clauses from the inductive invariant and observing a speed-up in the verification process.

In this paper, we will also use the side-loading technique to load the clauses from the inductive invariants generated from proving helpers with the IC3/PDR algorithm. In our method, we expect the convergence of proving the target property can benefit from the clauses associated with the helpers.

%% file: hls-inv/motivation.tex
\section{Motivation}\label{sec:motivation}

\begin{figure}[htbp]
     \centering
     \includegraphics[width=0.95\linewidth]{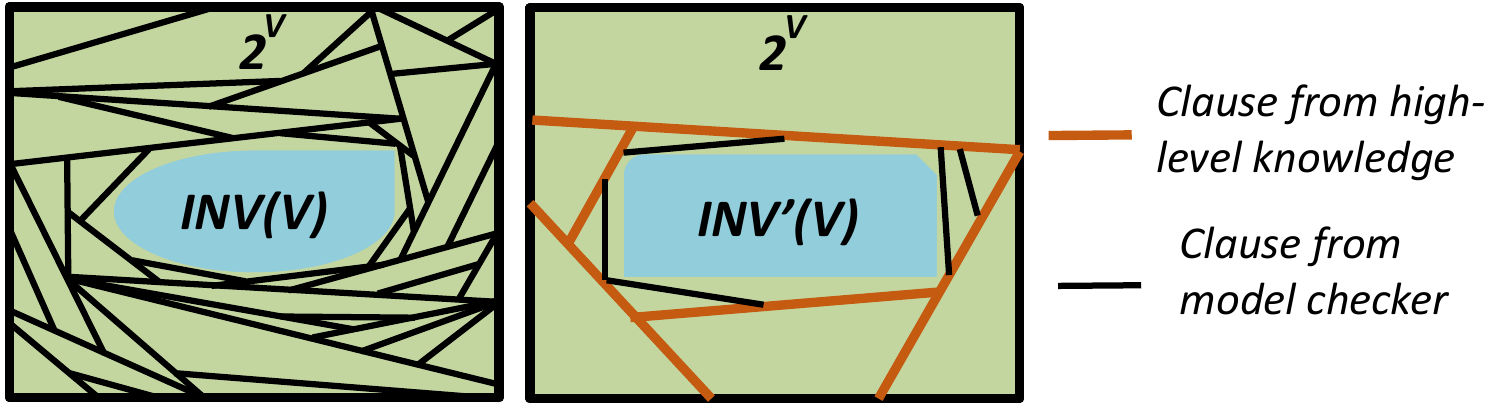}
     \vspace{-0.5\baselineskip}
     \caption{Comparison of finding an inductive invariant without or with constraints obtained from high-level design insights.}
     \label{fig:ic3_motivation} 
     \vspace{-0.7\baselineskip}
 \end{figure}

The verification process of the IC3/PDR algorithm fundamentally revolves around identifying an inductive invariant $\mathit{INV}(V)$ that effectively distinguishes reachable states from undesirable states. As previously discussed, the complexity of a design can significantly increase the time required to discover an inductive invariant within a reasonable timeframe. This challenge can be likened to sculpting a massive statue from a large block of stone using a small chisel. This analogy is depicted in the left rectangle of Figure \ref{fig:ic3_motivation}, where the black lines represent the clauses generated by the model checker, which constrain the state space $2^V$. \looseness=-1

To mitigate this issue, incorporating constraints based on high-level knowledge into the search space beforehand is akin to an experienced sculptor using a power cutter to remove unnecessary portions of the stone initially. This approach reduces the space required to identify the inductive invariant, as illustrated by the right rectangle in Figure \ref{fig:ic3_motivation}, where the brown lines eliminate a substantial portion of the state space, thereby significantly alleviating the burden on model checkers.

\begin{figure}[htb]
\centering
\vspace{-0.5\baselineskip}
\begin{subfigure}{0.50\linewidth}
    \begin{lstlisting}[linewidth=0.7\linewidth, captionpos=b,frame=none,numbers=none]
#define N 1000
void top(int a[N],
    int b[N],int c[N])
    for (int i=0;i<N;i++)
#pragma HLS pipeline
        c[i]=a[i]+b[i];
    \end{lstlisting}
\end{subfigure}
\hfill
\begin{subfigure}{0.42\linewidth}
\includegraphics[width=\linewidth]{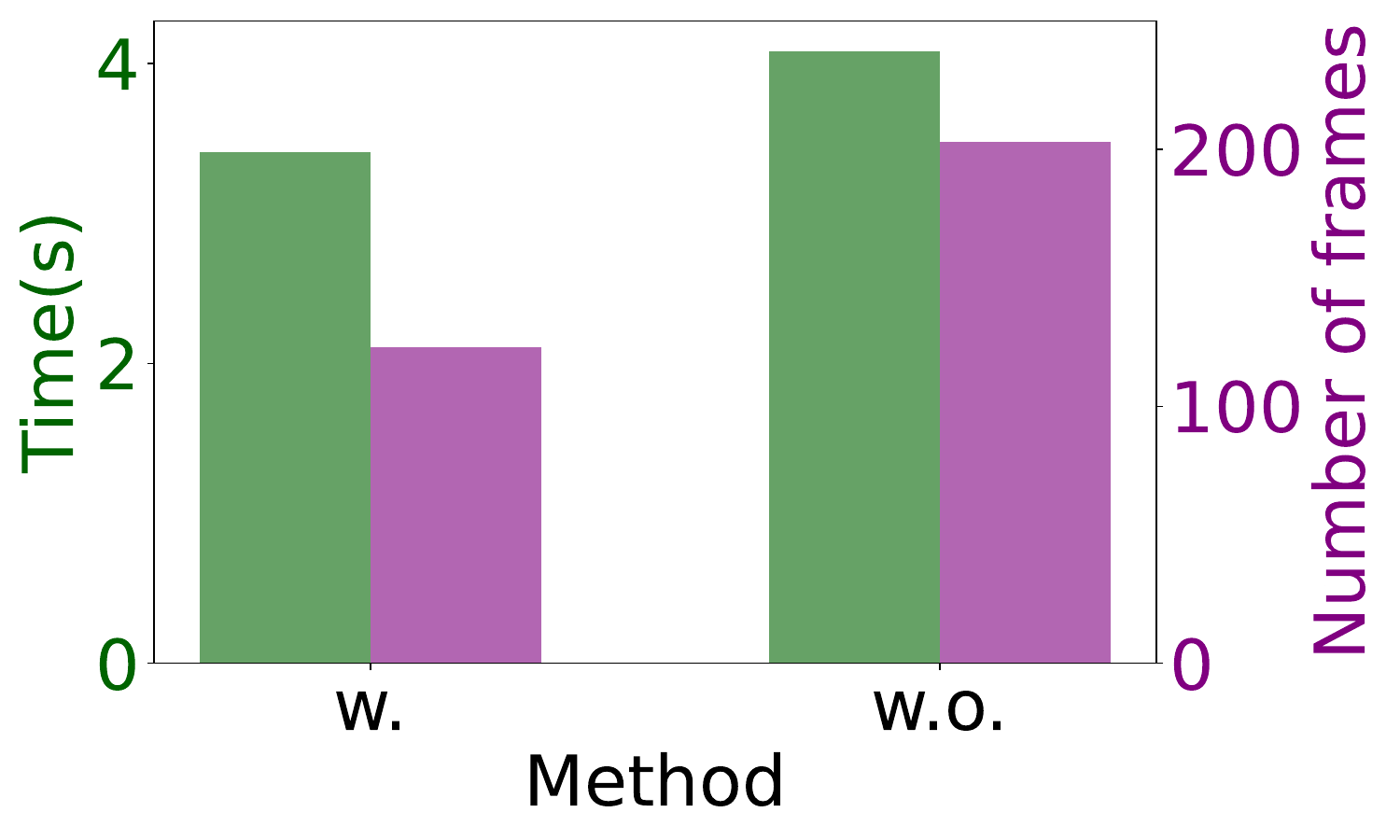}
\end{subfigure}
\vspace{-0.5\baselineskip}
\caption{Comparison of verification performance  with and without Helper on the vector addition example.}
\vspace{-0.5\baselineskip}
\label{fig:mot_s}
\end{figure}

Below is a motivating example. Consider a simple vector adder written in C++, as shown in the left part of Figure \ref{fig:mot_s}. Our goal is to find an execution trace that drives \texttt{ap\_done}, which indicates the hardware finishes execution, to 1 (PDR returns a witness trace if reachable). However, due to the requirement of thousands of state transitions to reach the completion state, the IC3/PDR algorithm must construct numerous frames to determine state reachability. 

By examining the source code and HLS reports produced from the HLS tool, it is evident that the upper bound of the loop variable is 1000. This high-level constraint can be incorporated into a helper assertion: \texttt{assert (i\_reg $\leq$ 1000)}. The proof of helper takes 0.07 seconds and results in an inductive invariant containing 4 clauses. These clauses are side-loaded to the proving process of the target property, enabling the IC3/PDR algorithm to find a trace leading to the \texttt{ap\_done} signal $1.19\times$ faster, with only 60.6\% of the frames constructed, as demonstrated in the right part of Figure \ref{fig:mot_s}.




%% file: hls-inv/method-HLSINV-overall.tex
\section{Proposed Framework: AutoINV}\label{sec:method}
\begin{figure}[htbp]
    \centering 
    \includegraphics[width=0.95\linewidth]{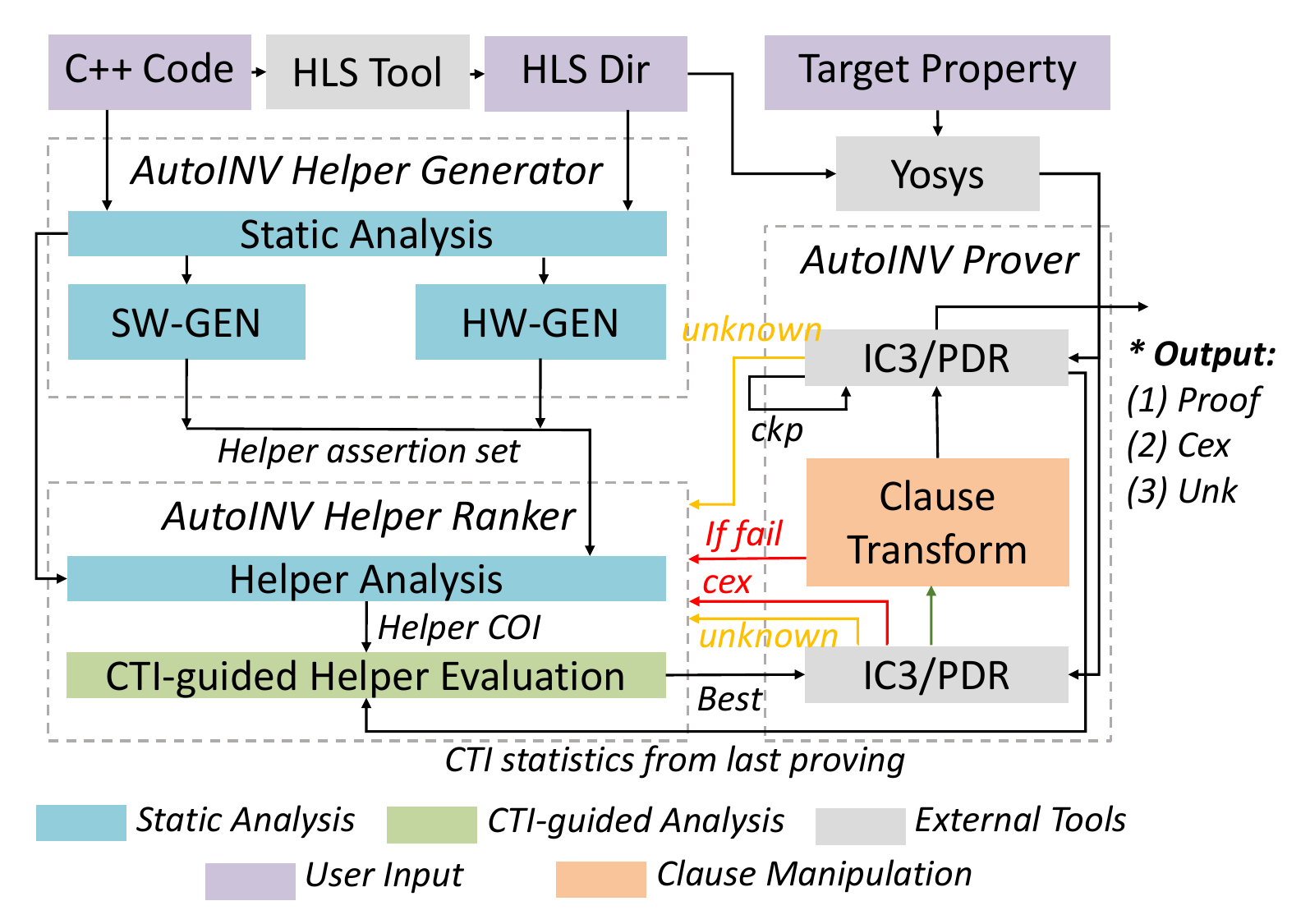} 
    \vspace{-0.5\baselineskip}
    \caption{AutoINV framework overview. SW-GEN and HW-GEN represent the software helper assertion generator and hardware helper assertion generator, respectively.
    \vspace{-1\baselineskip}
} 
    \label{fig:overview} 
\end{figure}

In this section, we provide a detailed explanation of the AutoINV framework. An overview of AutoINV is depicted in Figure \ref{fig:overview}. The framework accepts an HLS project directory and a user-defined target property written in \textit{SystemVerilog Assertion (SVA)} as inputs. The output of AutoINV is a proof or a counterexample trace violating the target property. When AutoINV runs out of time, the output will be unknown. The HLS project directory is required to contain both the HLS compilation report and the compiled HLS design in RTL. As previously introduced, AutoINV contains three parts: AutoINV Helper Generator generates helper assertions from analysis of RTL codes and high-level information of the design in HLS reports. AutoINV Helper Ranker ranks the helper assertions by importance according to the proving information from past trials. AutoINV Prover iteratively picks up the most significant helper assertion. It first checks the helper's validity. Then, it side-loads the clauses derived from the helper checking into the proving procedure of the target property. \looseness=-1
 
%% file: hls-inv/method-HLSINV-helper.tex
\subsection{AutoINV Helper Generator}

In the AutoINV Helper Generator, we identify and extract helper properties from the characteristic design patterns of HLS-generated hardware, with particular emphasis on control signals. These control signals exhibit highly consistent patterns and are typically encoded with less optimization than manually crafted RTL codes. In this work, we focus on identifying and leveraging these recurring signal patterns to extract effective helper properties.

The extraction process employs a pattern recognition approach that analyzes control signal relationships within the generated hardware. This approach does not necessitate perfect correctness in helper generation; when a proposed helper fails verification, it is eliminated, allowing AutoINV to proceed with alternative helper candidates without compromising overall verification correctness.

The AutoINV Helper Generator primarily focuses on two principal categories of properties as potential helpers: hardware-related properties, which examine structural characteristics of the implementation, and software-related properties, which derive from the behavioral specifications of the original algorithm.


\textbf{Hardware-related helpers} are generated from concise yet expressive facts in hardware using a rule-based design behavior characterization method. In this method, we mainly explore the behavioral pattern of three common design components in HLS designs: FIFO, FSM, and Pipeline.  

(A)  \textbf{FIFO}: The control part of FIFO in HLS designs is implemented with three groups of state registers: two indicating whether the FIFO is full or empty, and one representing the pointer to the data part. The data storage is implemented as either a shift register or a memory. We use the following rules to generate helpers regarding FIFO: When the pointer inside FIFO is not at its reset value, the FIFO is not empty;  When the pointer is not equal to the FIFO depth parameter, the FIFO is not full; the pointer of data storage should not exceed the FIFO depth; The FIFO should not be empty and full at the same time. \looseness=-1 

(B) \textbf{FSM}: For the sake of speed and power consumption of the generated designs, usually the FSM registers of each module inside HLS designs are one-hot encoded. 
As a result, the helper from FSM can be written as one-hot encoding,
where $FSM_i$ represent the $i$-th bit in the FSM registers.
Since FSM is the pump that drives the behavior of other modules, the clauses in the IC3/PDR algorithm are likely to contain literals referring to the FSM registers. The one-hot constraints for FSM could remind the IC3/PDR algorithm of the one-hot state transition constraints of FSM.


(C) \textbf{Pipeline}: In HLS designs, pipeline is a common structure to parallelize software loops in hardware execution \cite{hls_dse_2019,fado2, comba}. In a pipeline, the loop body in software is divided into several stages. During the execution, each stage processes a different part of the loop iteration, allowing for parallel execution and improved performance. The execution of each stage in HLS loops is controlled by two registers: one represents whether the stage is enabled, and another stores the calculated exiting condition for the stage. 

\begin{figure}[tbp]
    \centering
    \includegraphics[width=0.95\linewidth]{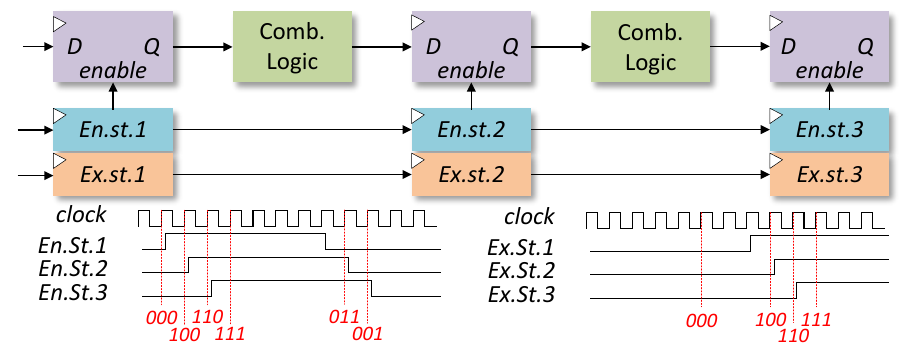}
    \vspace{-0.5\baselineskip}
    \caption{A simplified diagram of pipeline behavior in HLS designs during one execution of a loop. \textit{En.st.}$i$ and \textit{Ex.st.}$i$ represent the control registers for enabling and the exit condition at stage $i$, respectively.}
    \label{fig:pipeline}
    \vspace{-0.5\baselineskip}
\end{figure}

Owing to the nature of loop execution, the stages in the pipeline are enabled sequentially, and the satisfaction of the exiting condition for each stage is propagated sequentially accordingly through the stages. If we draw the behavior pattern of those stage control registers using a waveform, Figure~\ref{fig:pipeline} illustrates a simplified diagram of three execution stages inside the pipeline, which is accompanied by the two sets of control registers that represent the execution status of each stage. The waveform at the bottom of Figure~\ref{fig:pipeline} shows all possible behavioral patterns of control registers among the three pipeline stages. We can thus summarize the following rules: For both sets of pipeline exiting conditions and stage enabling registers, at most one register can be flipped in one cycle. Meanwhile, the register that is going to be flipped must follow a monotonic increasing or decreasing order. \looseness=-1

\begin{figure}[htbp!]
    \centering
    \begin{subfigure}[t]{0.95\linewidth}
        \centering
        \begin{subfigure}[h]{0.48\linewidth}
        \begin{lstlisting}[captionpos=b,frame=none,numbers=none]
void top(...) {
    int a = b * c;
    int d = a + b; }
        \end{lstlisting}            
        \end{subfigure}
        \hfill
        \begin{subfigure}[h]{0.48\linewidth}
            \includegraphics[width=0.9\linewidth]{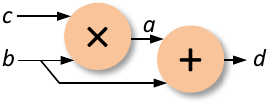}         
        \end{subfigure}
        \vspace{-0.5\baselineskip}
        \caption{Instruction execution order reflected in hardware implementation.}
        \label{fig:exe_order_instr}
    \end{subfigure}%
    
    \begin{subfigure}[t]{\linewidth}
        \centering
        \begin{subfigure}[h]{0.42\linewidth}
        \begin{lstlisting}[captionpos=b,frame=none,numbers=none]
void top(...) {
    int m[N];
    TASK1: for (...) {
#pragma HLS pipeline 
    /* compute and 
       write to m */}
    TASK2: for (...) {
#pragma HLS pipeline 
    /* read from m
       and compute */}}
        \end{lstlisting}            
        \end{subfigure}
        \hfill
        \begin{subfigure}[h]{0.55\linewidth}
            \includegraphics[width=\linewidth]{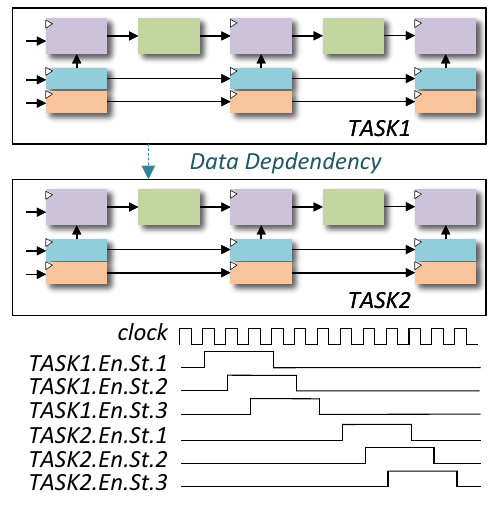}         
        \end{subfigure}
        \vspace{-0.5\baselineskip}
        \caption{Loop execution order reflected in hardware implementation.}
        \label{fig:exe_order_loop}
    \end{subfigure}

        \begin{subfigure}[t]{\linewidth}
        \centering
        \begin{subfigure}[h]{0.31\linewidth}
            \begin{subfigure}[h]{\linewidth}
            \begin{lstlisting}[captionpos=b,frame=none,numbers=none]
void top(...) {
    FUNC1(...);
    FUNC2(...);
    FUNC3(...);}
            \end{lstlisting}      
            \end{subfigure}
            \begin{subfigure}[h]{\linewidth}
                \centering
                \includegraphics[width=1.2\linewidth]{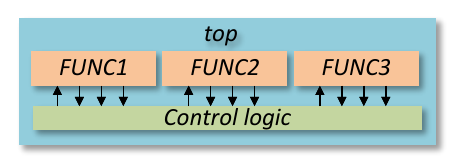}
            \end{subfigure}
            \vfill
        \end{subfigure}
        \hfill
        \begin{subfigure}[h]{0.60\linewidth}
            \centering
            \includegraphics[width=\linewidth]{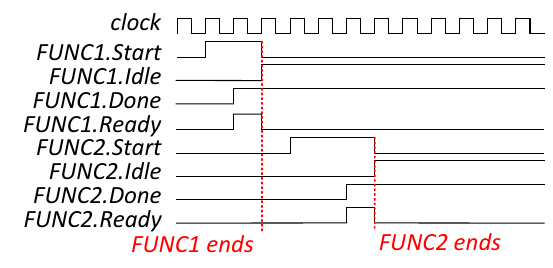}         
        \end{subfigure}
        \caption{Function execution order reflected in hardware implementation.}
        \label{fig:exe_order_func}
    \end{subfigure}
    \vspace{-0.5\baselineskip}
    \caption{Execution order reflected in hardware implementation.}
\end{figure}

\textbf{Software-related helpers} are mostly extracted from the software execution order defined in the input C++ code. Although HLS tools enable parallelization by mapping the input sequential software to hardware, certain execution orders should be kept owing to the inherent dependencies. Such an execution order can be reflected in the behavioral pattern of control signals. However, not all execution orders can be converted to helpers that can be mined. For example, the execution order inside a loop body that contains an addition followed by a multiplication is already guaranteed by the structure of the circuit: the output of the multiplier is connected to the input of the adder, shown in Figure \ref{fig:exe_order_instr}. In such a case, no control signals are involved, and the information of such an order cannot constrain the state space of the design. As a result, we only consider the execution order of loops and functions that are directly controlled by registers on the control path. In this work, we mine helpers from the coarse-grained execution order in HLS designs, namely loop-level helpers and function-level helpers.\looseness=-1

\textbf{Loop-level helpers} are helpers generated from the execution order of loops. Figure \ref{fig:exe_order_loop} illustrates a simple example: the input of instructions in loop \texttt{TASK2} is dependent on the results of loop \texttt{TASK1}, and both loops are pipelined. From the waveform of control registers for each loop, we can observe that the stage enable register in \texttt{TASK2} will not be set until the corresponding enable register in \texttt{TASK1} is reset, indicating that loop \texttt{TASK1} finishes the execution. For general cases, if any pair of loops are executed one after the other, we can derive a logic implication between the control signals:  If \texttt{TASK1} execute before \texttt{TASK2}, the set of enable registers in \texttt{TASK2} implies the reset of enable registers of \texttt{TASK1}. Similarly, for the registers that store exit condition for those loops, such an implication can be derived as well. \looseness=-1


\textbf{Function-level helpers} are helpers generated from the execution order of functions. These software functions are implemented as hardware modules in the generated RTL code from HLS tools. The function-level helpers are extracted from the block-level control interface functions. In this work, we consider a typical block-level control interface in HLS designs. It consists of four signals: (1) an input start signal that starts the execution of a module  (2) an output done signal indicating that the module finishes execution, (3) an output idle signal indicating that the module is currently idle (4) an output ready signal indicating that the module is ready to receive another group of data for processing. Similar to the execution order of loops, when a function is executed before another function, the block-level control interface will present a constrained behavior regarding the execution order. A general behavior pattern of such a control interface influenced by the execution order is shown in Figure \ref{fig:exe_order_func}. In this case, the function \texttt{FUNC1} is executed before the function \texttt{FUNC2}; the block-level control interface shows the following patterns in the waveform: the change of block-level control signal value of \texttt{FUNC2} is later than that of \texttt{FUNC1} and generally follows the same pattern. \looseness=-1 


\textbf{Loop variable helpers} are helpers generated when a loop has a fixed number of trip-count and is reflected directly from the loop variable. Such a constraint, although unrelated to the execution order, is an important part of the helper set. $R_i$ is the set of possible values of the loop variable in the $i$-th loop in the design. \looseness=-1

%% file: hls-inv/method-HLSINV-ranker.tex
\subsection{AutoINV Helper Ranker}\label{sec:ranker}

\begin{figure}[htbp]
\centering
\vspace{-0.5\baselineskip}
\includegraphics[width=0.95\linewidth]{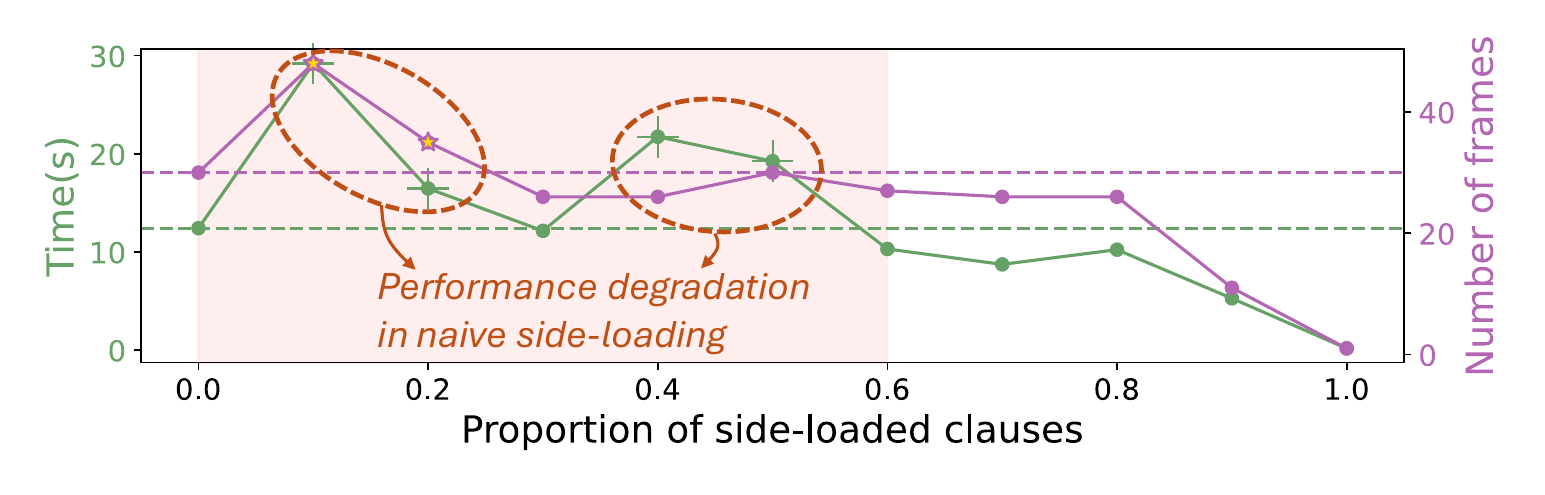}
\vspace{-0.5\baselineskip}
\caption{Performance degradation in naive clause side-loading.}
\label{fig:ranker_mot}
\vspace{-0.5\baselineskip}
\end{figure}

In our method, we use clause side-loading to accelerate model checking. However, not all clauses side-loaded into the proof engine lead to a quicker convergence of the IC3/PDR algorithm. Some unrelated side-loaded clauses potentially lead the algorithm in an undesired direction or cause extra costs when propagating them to higher levels. For example, in a simple equivalence checking task of different implementations of bit-wise XOR operation for two bit vectors, side-loading clauses naively without proper filtering may cause performance degradation even if the clauses are from the inductive invariant of exactly the same property. This is shown by our experimental results in Figure~\ref{fig:ranker_mot}. When side-loading  $10\%$, $20\%$ or $40\%$ clauses from inductive invariants, the checking time for the target property and the number of frames constructed for the IC3/PDR algorithm are actually worse than the case without side-loading.  What's more, compiling and verifying hundreds of helpers formally with the IC3/PDR algorithm could also be costly. Therefore, it is necessary to filter the set of helper assertions and only keep a subset of helpers beneficial to the proving process of the target property. However, as is mentioned in Section~\ref{sec:intro}, it is hard to select an optimal subset of helper assertion candidates produced by the AutoINV Helper Generator owing to an exponential number of possible combinations. We propose to derive a beneficial subset of helper assertions using preliminary trials when proving the target assertion. Specifically, we extract internal prover information related to \textit{counterexample-to-induction (CTI)} (detailed later) to evaluate the helpers. 
Such information in those trials reveals what the IC3/PDR algorithm is struggling with and is, therefore, meaningful for helper evaluation.
Based on this information, we construct an evaluation metric to rank the helper assertions by their potential benefits.

\begin{figure}
    \centering
    \vspace{-1.5\baselineskip}
    \includegraphics[width=0.9\linewidth]{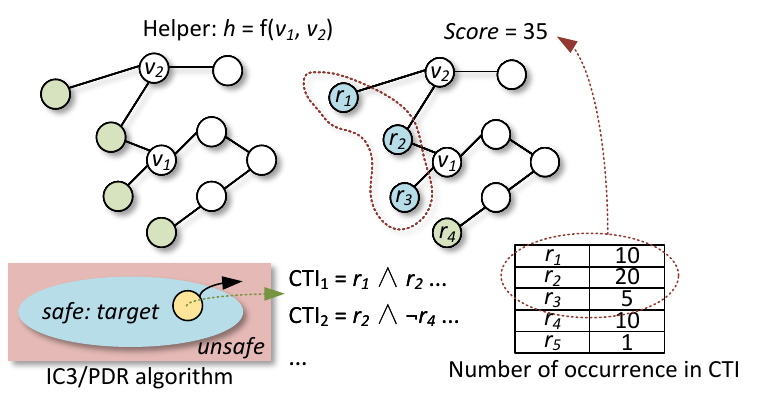}
    \caption{The workflow of AutoINV Helper Ranker.}
    \vspace{-0.5\baselineskip}
    \label{fig:ranker_work}
    \vspace{-0.5\baselineskip}
\end{figure}

In the bottom left of Figure~\ref{fig:overview}, AutoINV Helper Ranker takes as input the helper assertion set from AutoINV Helper Generator and the proving information from the last timeout trial. It provides a set of helper assertions with top scores once the downstream AutoINV Prover is asking for. AutoINV Helper Ranker evaluates a helper assertion in two steps: first, it utilizes the static information in the RTL code to analyze the helper assertion under evaluation; second, it employs the dynamic information from the last timeout proving trial on the target property to rank the priority of using the helper assertions. The details are shown in Figure~\ref{fig:ranker_work}.  

In the first step, we find the set of registers and inputs that are located in the \textit{transitive fan-in (TFI)} cone of variables in the assertion $h_i$. In the netlist of the circuit under verification, TFI of a variable represents the set of nodes from which there exists a path in the netlist from the node to the variable. Subsequently, we collect all registers and inputs found in the TFI cone as the set $\mathcal{R}_{h_i}$.




After AutoINV Helper Ranker derives $\mathcal{R}_{h_i}$, it then evaluates the benefit of the helper to the upcoming proving process based on the proving information of the previous trial. Here, we perform the evaluation based on CTI. In IC3/PDR, a CTI (counterexample to induction) is a cube that falsifies the inductiveness of a candidate invariant at some frame by exhibiting a predecessor that can transition to a violating state.
When a CTI is encountered, the IC3/PDR algorithm will start refining existing frames to block the CTI. 
As inductive invariants are generated piece by piece by blocking the CTIs, we center our evaluation method around CTIs and try to pick out helpers that possibly best assist in blocking CTIs. In AutoINV Helper Ranker, we use the number of occurrences of each variable in all CTIs encountered in the last proving attempt to calculate the score of helper $h_i$, denoted as $M_{h_i}$. Such a number of occurrences reflects the potential benefit of using a helper to constrain the variable. We sum these numbers of occurrences to get the score $M_{h_i}$. In this equation, $f_v(u)$ represents the frequency of the variable associated with node $u$ that appears in all CTIs in the last trial. \looseness=-1



%% file: hls-inv/method-HLSINV-prover.tex
\subsection{AutoINV Prover}\label{sec:hls-inv-prover}

\input{hls-inv/big_alg.tex}

AutoINV Prover is shown in the right part of Figure \ref{fig:overview}. It iteratively takes helpers and outputs the checking result of the target property. The core of AutoINV Prover is a proof engine implementing the IC3/PDR algorithm with checkpoint functionality, and it is responsible for proving helper assertions and the target property. The algorithm takes RTL design and the target property in \textit{and-inverter graph (AIG)}. The behavior of AutoINV Prover is defined by hyperparameters including the number of helper assertions taken from AutoINV Helper Ranker each time, denoted as $N_h$; the Timeout limit of the initial proving trial $T_0$; The growth rate of proving timeout limit $\alpha$; and the Timeout limit of proving helper assertion $T_h$. Details are provided in Algorithm \ref{alg:prover}.\looseness=-1

 In Algorithm \ref{alg:prover}, AutoINV Prover first launches an initial trial of prove engine with timeout limit $T_0$, as is shown in Line \ref{line:init_pdr}. Function $\text{pdr}(\cdot)$ is a single call of IC3/PDR prove engine, function arguments includes AIG of design and the target property under verification, timeout limit, side-loaded clauses, and checkpoint of the previous timeout trial. Checkpoint records clauses in the frames that are already constructed each time $\text{pdr}(\cdot)$ finishes the construction of a frame. The clauses learned from helper assertions are loaded onto the first frame. The IC3/PDR algorithm will attempt to push the side-loaded clause to higher frames in the propagation stage. If the target property is simple enough to derive a result within time $T_0$, AutoINV can directly return the proving result without any overhead of generating and evaluating the helper assertions. If the initial trial fails to derive any result within the timeout limit, the IC3/PDR algorithm will terminate once the current frame finishes construction. A checkpoint is dumped together with other related proving information. \looseness=-1

 After the initial trial, the algorithm reaches the main loop from line \ref{line:main_loop_start} to line \ref{line:main_loop_end}. Inside the $j$-th iteration, we first pick up the $N_h$ helper assertions with highest scores as $\mathbf{H}_j = \{h_{j,1}, h_{j,2}, \cdots, h_{j,N_h}\}$ from AutoINV Helper Ranker (Line \ref{line:helper_rank}). Yosys is used to compile every $h_{j,k}$ in $\mathbf{H}_j$ into AIG format with the design under verification. Line \ref{line:helper_compile} compiles the helper assertion into the corresponding AIG $A_{j,k}$. Next, in Line \ref{line:helper_pdr}, AutoINV Prover calls $\text{pdr}(\cdot)$ to verify the helper assertion within time $T_h$. If the $\text{pdr}(\cdot)$ function returns the helper to be incorrect or unknown within time $T_h$, it is abandoned and the AutoINV Helper Ranker will provide another set of promising helpers (Line \ref{line:helper_ask}). Once $h_{j,k}$ is proved, we can get an inductive invariant $I_{h_{j,k}}$ (Line \ref{line:get_ind_helper}), which contains a set of clauses learned from $h_{j,k}$. 

Before feeding the learned clauses into the $\text{pdr}(\cdot)$ function, it is required to pass a translation phase since the AIG that was used to prove $h_{j,k}$ is different from the AIG used to prove the target assertion. It is implemented in Function $\text{Translate}(\cdot)$ via variable name mapping. The translated inductive invariants are marked as $C_{h_{j,k}}$ (Line \ref{line:trans}). The set of clauses $C_{h_{j,k}}$ is later merged into the set of learned clauses of the current iteration (Line \ref{line:cup}). \looseness=-1

Line \ref{line:main_pdr} invokes the function $\text{pdr}(\cdot)$ to initiate a new attempt at proving the target property. This process leverages the checkpoint from the previous proving attempt and the learned clauses from $\mathbf{H_j}$. If this attempt exceeds the allotted time and the current frame completes its construction, the process is terminated, and a checkpoint is saved. Subsequently, the proving information is provided to the AutoINV Helper Ranker to select more promising helper assertions. Line \ref{line:timeout_update} adjusts the timeout limit for each trial by a factor of $\alpha$. It is important to note that once the timeout limit is reached in a proving trial, the IC3/PDR algorithm will not complete until the current frame finishes construction.

%% file: hls-inv/big_alg.tex
\begin{algorithm}[tbp]
\myalgsize
\caption{AutoINV Prover}
\label{alg:prover}
\begin{algorithmic}[1]
\State \textbf{Input:} AutoINV Helper Ranker object, the design under verification $A_t$ in AIG. 
\State \textbf{Parameters:} $\alpha$, $T_0$, $T_h$, $N_h$
\State \textbf{Output:} The proving result of the target property

\State $T \gets T_0; ckp \gets []$
\State $pdr\_result \gets \text{pdr}(A_t, T, \emptyset, ckp)$ \label{line:init_pdr}
\If {$pdr\_result.\text{is\_proved()} \vee pdr\_result.\text{is\_cex()}$ }
        \State \Return pdr\_result
\EndIf

\While{True} \label{line:main_loop_start}
    \State $\mathbf{H}_j \gets \text{ranker.get\_best\_helpers}(pdr\_result, N_h)$  \label{line:helper_rank}
    \State $C_j \gets \emptyset$
    \While{$\mathbf{H}_j \neq \emptyset$}
         \State $h_{j,k}$ $\gets$ an element in $\mathbf{H}_j$
         \State $\mathbf{H}_j \gets \mathbf{H}_j \setminus h_{j,k}$
         \State $A_{j,k}\gets \text{Compile}(h_{j,k}, A_t)$ \label{line:helper_compile}
         \State $pdr\_result\_helper \gets \text{pdr}(A_{j,k}, T_h, \emptyset, [])$ \label{line:helper_pdr}
         \If {not $pdr\_result\_helper.\text{is\_proved()}$}%
            \State $\mathbf{H}_j \gets \mathbf{H}_j\cup \text{ranker.get\_best\_helpers}(pdr\_result, 1)$ \label{line:helper_ask}
         \Else
         \State $I_{h_{j,k}} \gets pdr\_result\_helper.\text{get\_ind\_inv()}$ \label{line:get_ind_helper}
         \State $C_{h_{j,k}} \gets \text{Translate}(I_{h_{j,k}}, A_{j,k}, A_t)$ \label{line:trans}
         \State $C_j \gets C_j \cup C_{h_{j,k}}$ \label{line:cup}
         \EndIf
    \EndWhile
    
    \State $pdr\_result \gets \text{pdr}(A_t, T, C_j, ckp)$ \label{line:main_pdr}
    \If {$pdr\_result.\text{is\_proved()} \vee pdr\_result.\text{is\_cex()}$ }
        \State \Return pdr\_result
    \EndIf
    \State $ckp \gets pdr\_result$.$\text{get\_checkpoint()}$
    \State $T \gets \alpha T$ \label{line:timeout_update}
\EndWhile \label{line:main_loop_end}
\end{algorithmic}
\end{algorithm}

%% file: hls-inv/results.tex
\section{Evaluation}\label{sec:res}
\subsection{Experimental Setup}
Our experiments are conducted on a server running the Ubuntu 20.04 system. It is equipped with an Intel Core(TM) i5-11400 CPU working at a frequency of $4.2$GHz and 128GB of memory. 
We take the open-source IC3/PDR algorithm implementation in  IC3Ref~\cite{ic3ref22} and use AMD Vitis 2020.2 as the HLS tool. The target frequency for HLS compilation is 100MHz, and the target FPGA chip is xc7z020clg484-2. Yosys0.9 is used to compile the Verilog designs with assertions to the AIG file that can be taken as the input of IC3Ref. 
For RTL design analysis, we use PyVerilog 1.3.0~\cite{Takamaeda:2015:ARC:Pyverilog}.

\begin{table}[h]
\centering
\setlength\tabcolsep{2.3pt}
\caption{The scale of benchmarks.}
\vspace{-0.5\baselineskip}
\label{tab:res_resource}
\begin{tabular}{l||lllll}
\hline
    & \textit{vac-e} & \textit{vsc-e} & \textit{mac-e} & \textit{autosa-mm-f}~\cite{wang2021autosa} & \textit{autosa-cnn-f}~\cite{wang2021autosa} \\ \hline \hline
AND & 1468 & 1528 & 1781 & 11164 & 20821 \\
FF  & 204  & 213  & 242  & 1388  & 2941 \\
LOC & 1834 & 1454 & 2524 & 15925 & 25086 \\ \hline
\end{tabular}
\vspace{-1\baselineskip}
\end{table}

\subsection{Benchmarks}

\input{hls-inv/big_result_table.tex}

To validate the effectiveness of AutoINV, we employ several designs with their respective target properties as benchmarks: \textit{vac-e}, \textit{vsc-e}, \textit{mac-e}, \textit{autosa-mm-f}, and \textit{autosa-cnn-f}. In the experiment, we will first tune the hyperparameters of the AutoINV Prover using \textit{vac-e} (Functional consistency of vector accumulators), \textit{vsc-e} (Functional consistency of vector accumulators), and \textit{mac-e} (Functional consistency of MAC elements), shown in Section \ref{sec:hyper-param}. Subsequently, we will use the optimal hyperparameter configuration to evaluate the AutoINV framework on \textit{autosa-mm-f} and \textit{autosa-cnn-f} (FIFOs have no block in AutoSA-MM and AutoSA-CNN), which are two large-scale, real-world HLS designs generated by the  AutoSA~\cite{wang2021autosa} framework. \looseness=-2

The resource consumption and RTL code size of the benchmarks are reported in Table~\ref{tab:res_resource}. The resource consumption is detailed in the first and second rows of the table, indicating the number of AND gates and flip-flops in the AIG that feed into the IC3/PDR algorithm. This reflects the scale of properties after abstraction. The third row presents the lines of code in each RTL design. From these statistics, it is evident that the benchmarks are too complex in code size to perform manual helper extraction.






\subsection{Effectiveness of AutoINV Helper Generator}

In this subsection, we show the effectiveness of AutoINV Helper Generator in the helper generation task. We compare our helper generation scheme against HARM~\cite{HARM9925689}, a flexible assertion generation framework, to illustrate its efficiency in generating helpers for the verification of the target property. Table~\ref{tab:assertion_generation} presents the comparison results on three benchmarks: \textit{vac-e}, \textit{vsc-e}, and \textit{mac-e}. Using the default configuration of HARM, assertions were generated with the template restricted to the format of implication: $\text{condition} \rightarrow \text{result}$. 
In the provided cases, HARM requires hours to generate a large number of assertions using a fine-tuned ranking mechanism. Although the assertions generated by HARM maintain a reasonable correctness rate, they fail to produce even a few useful assertions for verification as helpers. If our framework attempts to generate and identify useful helpers from HARM, it will incur significant CPU time to generate a set of helper assertions, and even more time to find a useful subset that potentially benefits the verification process of the target property. In contrast, AutoINV can generate a relatively small but meaningful set of helpful assertions in a short time. The results demonstrate that the AutoINV helper generator is effective for helper generation in verification tasks on HLS designs and can quickly identify helpers that constrain the state space.


\begin{table}[htb]
\centering
\vspace{-0.5\baselineskip}
\caption{The comparison on assertion generation efficiency.}
\label{tab:assertion_generation}
\vspace{-0.5\baselineskip}
    \setlength\tabcolsep{2.5pt}
    \resizebox{\linewidth}{!}{
    \begin{tabular}{l||lll|l||lll|l}
    \hline
              & \multicolumn{4}{l||}{HARM~\cite{HARM9925689}}         & \multicolumn{4}{l}{AutoINV Helper Generator}    \\ \hline \hline
    Benchmark & Time(s)$^a$ & NUM$^b$ & CR$^c$$^e$ & UR$^d$$^e$ & Time(s) & NUM & CR & UR \\ \hline
    \textit{vac-e} & 5147.57 & 266273 & 100.0\% & 0.0\% & 3.02 & 83 & 75.9\% & \mygood{\textbf{73.5\%}}       \\ \hline
    \textit{vsc-e} & 3682.66 & 197749 & 44.7\%  & 0.2\% & 1.73 & 51 & 74.5\% & \mygood{\textbf{70.6\%}}     \\ \hline
    \textit{mac-e} & 4981.16 & 328498 & 79.9\%  & 0.2\% & 3.23 & 89 & 76.4\% & \mygood{\textbf{74.1\%}}      \\ \hline
    \multicolumn{9}{l}{$^a$ The CPU time cost for generating assertions. $^b$ Number of assertions.} \\
    \multicolumn{9}{l}{$^c$ The correctness rate (CR) of the top 1000 ranked assertions checked formally.} \\
    \multicolumn{9}{l}{$^d$ The useful rate (UR) of the top 1000 ranked assertions can produce clauses.} \\
    \multicolumn{9}{l}{$^e$ Both CR and UR are evaluated at a time frame limit of 10 seconds.}
    \end{tabular}
    }
\vspace{-1.5\baselineskip}
\end{table}

\subsection{Hyperparameter Configuration Evaluation}\label{sec:hyper-param}

In this subsection, we evaluate the effectiveness of AutoINV under different hyperparameter configurations on the AutoINV Prover. The hyper-parameters include $N_h$, $T_0$, and $\alpha$, as introduced in Section~\ref{sec:hls-inv-prover}. We limit $T_h$ to 10 seconds. The configurations are shown in Table~\ref{tab:config}. In our experiments, five hyperparameter configurations are used. The ``Raw'' configuration in the table refers to the vanilla IC3/PDR algorithm without any side-loading.

\begin{table}[b]
\centering
\vspace{-1\baselineskip}
\caption{Configurations of hyperparameters on AutoINV Prover.}
\vspace{-0.5\baselineskip}
\label{tab:config}

\setlength\tabcolsep{3pt}
\begin{tabular}{l||lll||llll||llll}
\hline
Config & $N_h$ & $T_0$  & $\alpha$ & Config & $N_h$ & $T_0$  & $\alpha$ & Config & $N_h$ & $T_0$  & $\alpha$  \\ \hline
Raw    & -     & $\inf$ & -        &  1     & 5     & 10     & 5        &  2     & 5     & 10     & 2       \\  
3      & 5     & 100    & 2        &  4     & 10    & 1      & 2        &  5     & 10    & 100    & 2      \\ 
\hline
\end{tabular}
\end{table}

During verification, every task was proven as expected. Table \ref{tab:ic3_pdr_res} provides results for \textit{vac-e}, \textit{vsc-e}, and \textit{mac-e} under different configurations. 
From the statistics, we observe that in \textit{vac-e}, AutoINV achieves a speed-up ranging from 1.19$\times$ to 1.57$\times$ compared to the vanilla IC3/PDR algorithm; in \textit{vsc-e}, AutoINV achieves a speed-up of 3.56$\times$ to 6.05$\times$; and in \textit{mac-e}, AutoINV achieves a speed-up of 2.08$\times$ to 3.14$\times$. These speed-up ratios demonstrate the effectiveness of our verification framework for formal verification tasks of HLS designs. \looseness=-1

Additionally, in most cases, the speed-up is accompanied by a reduced number of SAT solver calls and CTIs encountered. For example, in \textit{vsc-e}, the number of SAT solver calls is reduced by 42.9\% to 64.9\%, and the CTIs encountered during verification are reduced by 51.5\% to 64.9\%. This indicates that our method can prune the state space early in the verification process and guide the IC3/PDR algorithm towards a better direction for searching for potential inductive invariants or counterexamples. \looseness=-1


Although there are some cases, such as configuration 1 and configuration 2, when verifying \textit{vac-e}, where more SAT queries or CTIs are required compared to the baseline, the total verification time is still reduced. This is due to the clauses added by AutoINV simplifying the SAT engine's solving process or better generalization of CTIs that require more SAT queries. Better generalization of CTIs is reflected by the reduction in the number of counterexamples to generalization (CTG) by 16.44\% and 4.34\%, respectively, despite the increase in CTIs in configuration 2.



In most cases, AutoINV enables the IC3/PDR algorithm to construct fewer frames to converge to the inductive invariant, demonstrating our framework's ability to guide the IC3/PDR algorithm towards a more radical approximation of the reachable states.

The average speed-up for each configuration is shown in the last column, indicating that AutoINV can achieve an average speed-up range from 2.11$\times$ to 2.47$\times$ regardless of the configurations. In the next subsection, we will use configuration 4, which has the highest speed-up of 2.47$\times$, to conduct verification tasks on large designs.

\subsection{Results on AutoSA designs}


This subsection presents the results of AutoINV on two AutoSA designs. We set the timeout bound for AutoINV to 12 hours. The hyperparameters for the AutoINV Prover are set to configuration 4, as concluded from Section \ref{sec:hyper-param}. Table \ref{tab:res_autosa} shows the results of the verification tasks: \textit{autosa-mm-f} and \textit{autosa-cnn-f}.\looseness=-2

In \textit{autosa-mm-f}, AutoINV achieves a 4.76$\times$ speed-up compared to the vanilla IC3/PDR algorithm without side-loading. With proper guidance from our methodology, the number of SAT queries is reduced by 83.0\%, and the number of CTIs encountered is reduced by 38.5\%. IC3/PDR produces a 35-cycle counterexample trace that drives the FIFO to the full state. This suggests that the FIFO length should be increased in the original design to avoid a potential FIFO block that could affect the overall execution speed.\looseness=-2

In \textit{autosa-cnn-f}, the target property is proven by AutoINV in 322.09 seconds, indicating that there is no execution trace that causes the FIFO to become full. However, the vanilla IC3/PDR algorithm fails to derive a proof within the timeout bound. \looseness=-1

\begin{table}[htb]
\centering
\vspace{-0.5\baselineskip}
\caption{IC3/PDR results on AutoSA benchmarks.}
\vspace{-0.5\baselineskip}
\setlength\tabcolsep{3.0pt}
\begin{tabular}{c||llllllc}
\hline 
Config  & Time(s) & SAT.Q & CTI & IT & CL & FR & Result \\ \hline \hline
       & \multicolumn{7}{c}{\textit{autosa-mm-f}~\cite{wang2021autosa}} \\ \hline

4      & 468.88  & 486608  & 8817  & 5 & 436 & 35 & Cex \\
Raw    & 2235.16 & 2858343 & 14346 & 1 & 0   & 35 & Cex \\ \hline
       & \multicolumn{7}{c}{\textit{autosa-cnn-f}~\cite{wang2021autosa}} \\ \hline
4      & 322.09  & 14454   & 254   & 4 & 240 & 38 &Proof\\
Raw    & O.O.T.  & - &-&-&-&-&Unknown\\ \hline
\end{tabular}
\vspace{-0.5\baselineskip}
\label{tab:res_autosa}
\vspace{-0.5\baselineskip}
\end{table}

%% file: hls-inv/big_result_table.tex
\begin{table*}[t]
\centering
\vspace{-0.5\baselineskip}
\caption{IC3/PDR results on benchmarks for hyper-parameter tuning on various configurations of AutoINV Prover.}
\vspace{-0.5\baselineskip}
\label{tab:ic3_pdr_res}
\setlength\tabcolsep{2.0pt}
\resizebox{\linewidth}{!}{

\begin{tabular}{c||lllll|ll||lllll|ll||lllll|ll||l}
\hline
        & \multicolumn{7}{l||}{\textit{vac-e}}                     & \multicolumn{7}{l||}{\textit{vsc-e}}                     & \multicolumn{7}{l||}{\textit{mac-e}} & AVG$^h$                     \\ \hline
Config  & SAT.Q$^a$  & CTI$^b$  & IT$^c$ & CL$^d$ & FR$^e$ & Time(s)$^f$      & Acc$^g$ &  SAT.Q & CTI & IT & CL & FR & Time(s) & Acc &  SAT.Q & CTI & IT & CL & FR & Time(s) & Acc & Acc \\ \hline \hline
Raw & 12111500 & 78817 & 1 & 0 & 42 & 8310.22 & \multicolumn{1}{c||}{-}
    & 23354152 & 71560 & 1 & 0 & 95 & 7704.33 & \multicolumn{1}{c||}{-}
    & 16317585 & 36303 & 1 & 0 & 106& 6239.64 & \multicolumn{1}{c||}{-}
    & \multicolumn{1}{c}{-}\\ \hline    
1       &  12991356   &  77729   &  5    &   210  &  50    & 6972.62          &  1.19$\times$  
        &  8731930    &  27299   &  4    &   185  &  64    & 1510.68          &  5.10$\times$ 
        &  10561985   &  33246   &  4    &   161  &  97    & 2190.29 &  2.84$\times$
        & 2.11$\times$\\
2       & 11620294 & 81727 & 9 & 183 & 38 & 6751.35 & 1.23$\times$ 
        & 9913628  & 31741 & 6 & 200 & 84 & 1620.94 & 4.75$\times$ 
        & 9586488  & 29785 & 7 & 248 & 90 & \mygood{\textbf{1984.17}} & \mygood{\textbf{3.14$\times$}}
        & 2.17$\times$\\
3       & 10249622 & 74325 & 6 & 155 & 38  & \mygood{\textbf{5288.30}} & \mygood{\textbf{1.57$\times$}}
        & 13327248 & 34672 & 5 & 192 & 122 & 2162.91          & 3.56$\times$
        & 11956644 & 30859 & 5 & 234 & 97  & 2996.11          & 2.08$\times$ 
        & 2.16$\times$ \\
4       & 10768763 & 76494 & 9 & 236 & 44 & 5470.70          & 1.51$\times$ 
        & 8270314  & 25082 & 7 & 213 & 97 & \mygood{\textbf{1271.50}} & \mygood{\textbf{6.05$\times$}}
        & 9825256  & 28152 & 9 & 295 & 85 & 2383.76          & 2.61$\times$
        & \mygood{\textbf{2.47$\times$}}\\
5       & 10016820 & 73685 & 6 & 221 & 36 & 5350.88 & 1.55$\times$
        & 10981527 & 29735 & 4 & 207 & 96 & 1850.83 & 4.16$\times$
        & 10067206 & 29840 & 5 & 262 & 86 & 2872.73 & 2.17$\times$
        & 2.23$\times$\\ \hline
\multicolumn{22}{l}{$^a$ The number of SAT solver calls in the verification process. $^b$ The number of CTIs encountered. $^c$ The number of iterations of the AutoINV Prover.} \\
\multicolumn{22}{l}{$^d$ The number of side-loaded clauses. $^e$ The number of frames constructed in the IC3/PDR algorithm. $^f$ The time usage to prove the target property.} \\
\multicolumn{22}{l}{$^g$ The speed-up ratio under different hyper-parameter configurations. $^h$ The average speed-up ratio of benchmarks provided.}
\end{tabular}
}
\vspace{-1.5\baselineskip}

\end{table*}

%% file: hls-inv/conclusion.tex
\section{Conclusion}\label{sec:conclusion}

In this paper, we propose AutoINV, a framework that guides formal verification of hardware properties in HLS designs. AutoINV generates helper-assertion candidates from common HLS patterns and dynamically prioritizes effective ones based on CTIs observed in timeout trials. It iteratively applies selected helpers until IC3/PDR proves the target property. Experiments on diverse HLS designs and properties demonstrate that AutoINV significantly improves verification speed. \looseness=-1


%% file: hls-inv.bib
@String{Computing = "Computing" }

@String{Computer = "{IEEE} Computer" }

@String{Springer = "Springer-Verlag" }

@inproceedings{Herklotz2021,
   author = {Yann Herklotz and Zewei Du and Nadesh Ramanathan and John Wickerson},
   doi = {10.1109/FCCM51124.2021.00034},
   journal = {Proceedings - 29th IEEE International Symposium on Field-Programmable Custom Computing Machines, FCCM 2021},
   title = {An Empirical Study of the Reliability of High-Level Synthesis Tools},
   year = {2021},
}

@article{Pundir2022,
   author = {Nitin Pundir and Sohrab Aftabjahani and Rosario Cammarota and Mark Tehranipoor and Farimah Farahmandi},
   doi = {10.1145/3492345},
   issn = {15504840},
   issue = {3},
   journal = {ACM Journal on Emerging Technologies in Computing Systems},
   title = {Analyzing Security Vulnerabilities Induced by High-level Synthesis},
   volume = {18},
   year = {2022},
}

@inproceedings{bradley2011sat,
  title={SAT-based model checking without unrolling},
  author={Bradley, Aaron R},
  booktitle={International Workshop on Verification, Model Checking, and Abstract Interpretation},
  pages={70--87},
  year={2011},
  organization={Springer}
}

@inproceedings{pdr-een2011efficient,
  title={Efficient implementation of property directed reachability},
  author={E{\'e}n, Niklas and Mishchenko, Alan and Brayton, Robert},
  booktitle={2011 Formal Methods in Computer-Aided Design (FMCAD)},
  pages={125--134},
  year={2011},
  organization={IEEE}
}

@inproceedings{kang2023lfps,
  title={Lfps: Learned formal proof strengthening for efficient hardware verification},
  author={Kang, Minwoo and Nova, Azade and Singh, Eshan and Bathini, Geetheeka Sharron and Viktorov, Yuriy},
  booktitle={2023 IEEE/ACM International Conference on Computer Aided Design (ICCAD)},
  pages={1--9},
  year={2023},
  organization={IEEE}
}

@inproceedings{hu2024deepic3,
  title={DeepIC3: Guiding IC3 Algorithms by Graph Neural Network Clause Prediction},
  author={Hu, Guangyu and Tang, Jianheng and Yu, Changyuan and Zhang, Wei and Zhang, Hongce},
  booktitle={2024 29th Asia and South Pacific Design Automation Conference (ASP-DAC)},
  pages={262--268},
  year={2024},
  organization={IEEE}
}

@inproceedings{Takamaeda:2015:ARC:Pyverilog,
title={Pyverilog: A Python-Based Hardware Design Processing Toolkit for Verilog HDL},
author={Takamaeda-Yamazaki, Shinya},
booktitle={Applied Reconfigurable Computing},
month={Apr},
year={2015},
pages={451-460},
volume={9040},
series={Lecture Notes in Computer Science},
publisher={Springer International Publishing},
doi={10.1007/978-3-319-16214-0_42},
url={http://dx.doi.org/10.1007/978-3-319-16214-0_42},
}

@misc{ic3ref22,
  author = {Bradley, A.},
  title = {An implementation of IC3},
  year = {2022},
  url = {https://github.com/arbrad/IC3ref}
}

@inproceedings{piccolboni2019kairos,
  title={Kairos: Incremental verification in high-level synthesis through latency-insensitive design},
  author={Piccolboni, Luca and Di Guglielmo, Giuseppe and Carloni, Luca P},
  booktitle={2019 Formal Methods in Computer Aided Design (FMCAD)},
  pages={105--109},
  year={2019},
  organization={IEEE}
}

@inproceedings{li2023se3,
  title={SE3: Sequential Equivalence Checking for Non-Cycle-Accurate Design Transformations},
  author={Li, You and Zhao, Guannan and He, Yunqi and Zhou, Hai},
  booktitle={2023 60th ACM/IEEE Design Automation Conference (DAC)},
  pages={1--6},
  year={2023},
  organization={IEEE}
}

@inproceedings{xu2023automaticdynopt,
  title={Automatic inductive invariant generation for scalable dataflow circuit verification},
  author={Xu, Jiahui and Josipovi{\'c}, Lana},
  booktitle={2023 IEEE/ACM International Conference on Computer Aided Design (ICCAD)},
  pages={1--9},
  year={2023},
  organization={IEEE}
}

@ARTICLE{HARM9925689,
  author={Germiniani, Samuele and Pravadelli, Graziano},
  journal={IEEE Transactions on Computer-Aided Design of Integrated Circuits and Systems}, 
  title={HARM: A Hint-Based Assertion Miner}, 
  year={2022},
  volume={41},
  number={11},
  pages={4277-4288},
  doi={10.1109/TCAD.2022.3197525}}

@INPROCEEDINGS{2010goldmine,
  author={Vasudevan, Shobha and Sheridan, David and Patel, Sanjay and Tcheng, David and Tuohy, Bill and Johnson, Daniel},
  booktitle={2010 Design, Automation \& Test in Europe Conference \& Exhibition (DATE 2010)}, 
  title={GoldMine: Automatic assertion generation using data mining and static analysis}, 
  year={2010},
  volume={},
  number={},
  pages={626-629},
  doi={10.1109/DATE.2010.5457129}}

@inproceedings{assertLLM,
author = {Yan, Zhiyuan and Fang, Wenji and Li, Mengming and Li, Min and Liu, Shang and Xie, Zhiyao and Zhang, Hongce},
title = {AssertLLM: Generating Hardware Verification Assertions from Design Specifications via Multi-LLMs},
year = {2025},
isbn = {9798400706356},
publisher = {Association for Computing Machinery},
address = {New York, NY, USA},
url = {https://doi.org/10.1145/3658617.3697756},
doi = {10.1145/3658617.3697756},
booktitle = {Proceedings of the 30th Asia and South Pacific Design Automation Conference},
pages = {614–621},
numpages = {8},
location = {Tokyo, Japan},
series = {ASPDAC '25}
}

@INPROCEEDINGS{goldmine_word,
  author={Liu, Lingyi and Lin, Chen-Hsuan and Vasudevan, Shobha},
  booktitle={2012 IEEE/ACM International Conference on Computer-Aided Design (ICCAD)}, 
  title={Word level feature discovery to enhance quality of assertion mining}, 
  year={2012},
  volume={},
  number={},
  pages={210-217},
  doi={}}

@INPROCEEDINGS{goldmine_memp,
  author={Liu, Lingyi and Sheridan, David and Athavale, Viraj and Vasudevan, Shobha},
  booktitle={Ninth ACM/IEEE International Conference on Formal Methods and Models for Codesign (MEMPCODE2011)}, 
  title={Automatic generation of assertions from system level design using data mining}, 
  year={2011},
  volume={},
  number={},
  pages={191-200},
  doi={10.1109/MEMCOD.2011.5970526}}

@inproceedings{iodine05,
author = {Hangal, Sudheendra and Chandra, Naveen and Narayanan, Sridhar and Chakravorty, Sandeep},
title = {IODINE: a tool to automatically infer dynamic invariants for hardware designs},
year = {2005},
isbn = {1595930582},
publisher = {Association for Computing Machinery},
address = {New York, NY, USA},
url = {https://doi.org/10.1145/1065579.1065786},
doi = {10.1145/1065579.1065786},
booktitle = {Proceedings of the 42nd Annual Design Automation Conference},
pages = {775–778},
numpages = {4},
location = {Anaheim, California, USA},
series = {DAC '05}
}

@INPROCEEDINGS{parallel2020FMCAD,
  author={Dureja, Rohit and Baumgartner, Jason and Kanzelman, Robert and Williams, Mark and Rozier, Kristin Y.},
  booktitle={2020 Formal Methods in Computer Aided Design (FMCAD)}, 
  title={Accelerating Parallel Verification via Complementary Property Partitioning and Strategy Exploration}, 
  year={2020},
  volume={},
  number={},
  pages={16-25},
  doi={10.34727/2020/isbn.978-3-85448-042-6_8}}

@INPROCEEDINGS{parallel20190FMCAD,
  author={Dureja, Rohit and Baumgartner, Jason and Ivrii, Alexander and Kanzelman, Robert and Rozier, Kristin Y.},
  booktitle={2019 Formal Methods in Computer Aided Design (FMCAD)}, 
  title={Boosting Verification Scalability via Structural Grouping and Semantic Partitioning of Properties}, 
  year={2019},
  volume={},
  number={},
  pages={1-9},
  doi={10.23919/FMCAD.2019.8894265}}

@INPROCEEDINGS{purse24date,
  author={Das, Sourav and Hazra, Aritra and Dasgupta, Pallab and Kundu, Sudipta and Jain, Himanshu},
  booktitle={2024 Design, Automation \& Test in Europe Conference \& Exhibition (DATE)}, 
  title={PURSE: Property Ordering Using Runtime Statistics for Efficient Multi - Property Verification}, 
  year={2024},
  volume={},
  number={},
  pages={1-6},
  doi={10.23919/DATE58400.2024.10546895}}

@article{FMHLS,
author = {Herklotz, Yann and Pollard, James D. and Ramanathan, Nadesh and Wickerson, John},
title = {Formal verification of high-level synthesis},
year = {2021},
issue_date = {October 2021},
publisher = {Association for Computing Machinery},
address = {New York, NY, USA},
volume = {5},
number = {OOPSLA},
url = {https://doi.org/10.1145/3485494},
doi = {10.1145/3485494},
journal = {Proc. ACM Program. Lang.},
month = oct,
articleno = {117},
numpages = {30}
}

@INPROCEEDINGS{FMHLS_inline,
  author={Pardalos, Michalis and Herklotz, Yann and Wickerson, John},
  booktitle={2022 IEEE 30th Annual International Symposium on Field-Programmable Custom Computing Machines (FCCM)}, 
  title={Resource Sharing for Verified High-Level Synthesis}, 
  year={2022},
  volume={},
  number={},
  pages={1-6},
  doi={10.1109/FCCM53951.2022.9786208}}

@ARTICLE{hls_dse_2019,
  author={Schafer, Benjamin Carrion and Wang, Zi},
  journal={IEEE Transactions on Computer-Aided Design of Integrated Circuits and Systems}, 
  title={High-Level Synthesis Design Space Exploration: Past, Present, and Future}, 
  year={2020},
  volume={39},
  number={10},
  pages={2628-2639},
  doi={10.1109/TCAD.2019.2943570}}

@article{fado2,
author = {Du, Linfeng and Liang, Tingyuan and Zhou, Xiaofeng and Ge, Jinming and Li, Shangkun and Sinha, Sharad and Zhao, Jieru and Xie, Zhiyao and Zhang, Wei},
title = {FADO: Floorplan-Aware Directive Optimization Based on Synthesis and Analytical Models for High-Level Synthesis Designs on Multi-Die FPGAs},
year = {2024},
issue_date = {September 2024},
publisher = {Association for Computing Machinery},
address = {New York, NY, USA},
volume = {17},
number = {3},
issn = {1936-7406},
url = {https://doi.org/10.1145/3653458},
doi = {10.1145/3653458},
journal = {ACM Trans. Reconfigurable Technol. Syst.},
month = sep,
articleno = {47},
numpages = {33}
}

@INPROCEEDINGS{comba,
  author={Zhao, Jieru and Feng, Liang and Sinha, Sharad and Zhang, Wei and Liang, Yun and He, Bingsheng},
  booktitle={2017 IEEE/ACM International Conference on Computer-Aided Design (ICCAD)}, 
  title={COMBA: A comprehensive model-based analysis framework for high level synthesis of real applications}, 
  year={2017},
  volume={},
  number={},
  pages={430-437},
  doi={10.1109/ICCAD.2017.8203809}}

@inproceedings{wang2021autosa,
  title={AutoSA: A Polyhedral Compiler for High-Performance Systolic Arrays on FPGA},
  author={Wang, Jie and Guo, Licheng and Cong, Jason},
  booktitle={Proceedings of the 2021 ACM/SIGDA International Symposium on Field-Programmable Gate Arrays},
  year={2021}
}
